\documentclass[
aps,prl,twocolumn,
superscriptaddress,%showpacs,%11pt,%12pt,
amsfonts,amssymb,amsmath,
twoside,
letterpaper,%a4paper,
bibnotes%,preprint
]{revtex4}

\pdfoutput=1

\usepackage{bm,color,mathrsfs,graphicx}

\begin{document}

\title{
Multi-domain ferroelectricity as a limiting factor for \\
voltage amplification in ferroelectric field-effect transistors
}

\author{A. Cano}
\email{cano@esrf.fr}
\affiliation{
European Synchrotron Radiation Facility, 6 rue Jules Horowitz, BP 220, 38043 Grenoble, France}

\author{D. Jim\'{e}nez }
\email{david.jimenez@uab.es}
\affiliation{
%\mbox{
Departament d'Enginyeria Electr\`{o}nica, Escola d'Enginyeria, 
Universitat Aut\`{o}noma de Barcelona, 08193 Bellaterra, Spain
%}
}

\date{\today}

\begin{abstract} 
We revise the possibility of having an amplified surface potential in ferroelectric field-effect transistors pointed out by S. Salahuddin and S. Datta [Nano Lett. {\bf 8}, 405 (2008)]. We show that the negative-capacitance regime that allows for such an amplification is actually bounded by the appearance of multi-domain ferroelectricity. This imposes a severe limit to the maximum step-up of the surface potential obtainable in the device.
We indicate new device design rules taking into account this scenario. 
\end{abstract}

\pacs{
%77.80.-e %Ferroelectricity and antiferroelectricity  
%77.55.+f %Dielectric thin films
%77.22.Ej %Polarization and depolarization  
}

\maketitle

\begin{figure}[b]
\includegraphics[width=.475\textwidth]{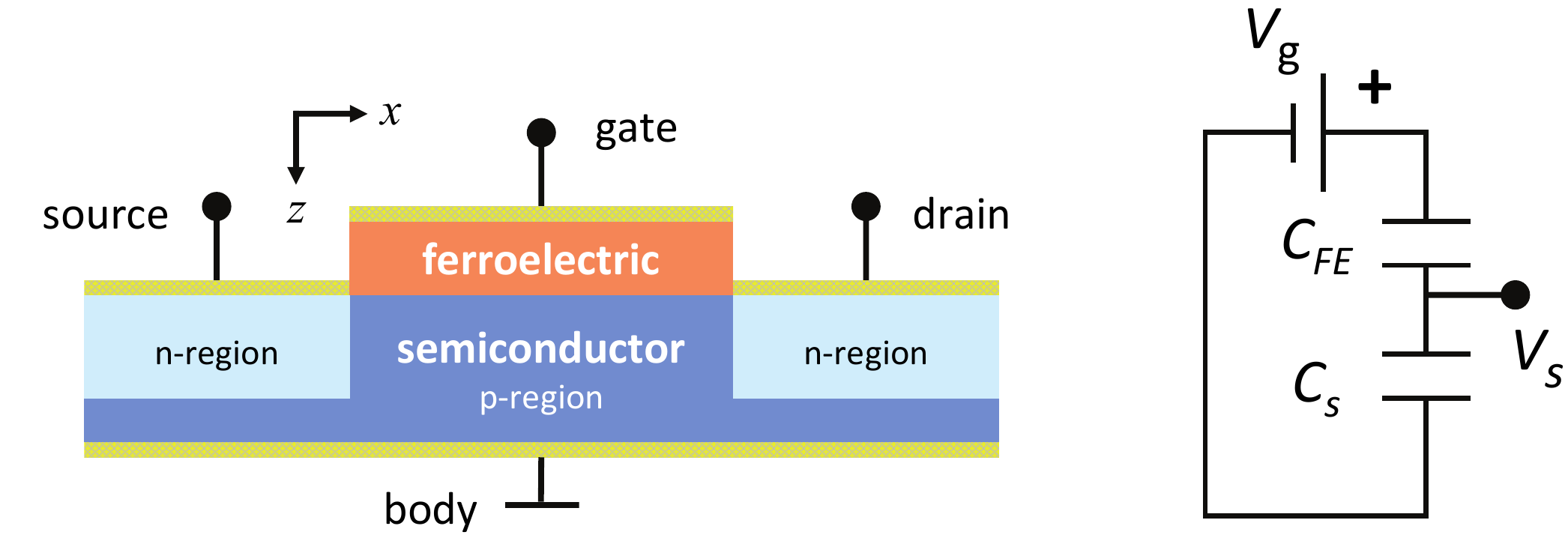}
\caption{Schematic cross section of a ferroelectric FET (left) and equivalent circuit for the metal/ferroelectric/semiconductor stack (right).}
\label{FE-FET}
\end{figure}

The operation of field-effect transistors (FETs) generates a heat whose dissipation imposes severe restrictions to the miniaturization of integrated circuits. 
The lowering of the FET operating voltage is therefore highly desirable, which has to be accompanied with a reduction of the threshold voltage to mantain performaces. This, however, implies the increse in the stand-by power since the inverse of the so-called subthreshold slope appears limited to 60 mV/decade at room temperature. At present, this is considered as an important roadblock for the transistor scaling down \cite{Roadmap}. 
Recently Salahuddin and Datta have suggested that the 60 mV/decade limit can be overcome 
in ferroelectric FETs as the sketched in Fig \ref{FE-FET} \cite{Datta08}. 
These FETs has a long history as canditates for nondestructive readout memory elements \cite{Dawber05}.
The idea of Salahuddin and Datta is to exploit a negative capacitance regime of the ferroelectric in which the surface potential of the semiconductor $V_s$ is up-converted and therefore the so-called body factor of the transistor $m \equiv (\partial V_s/\partial V_g)^{-1}$, where $V_g$ is the gate potential, becomes smaller than one.

\begin{figure*}[t]
\includegraphics[width=.85\textwidth]{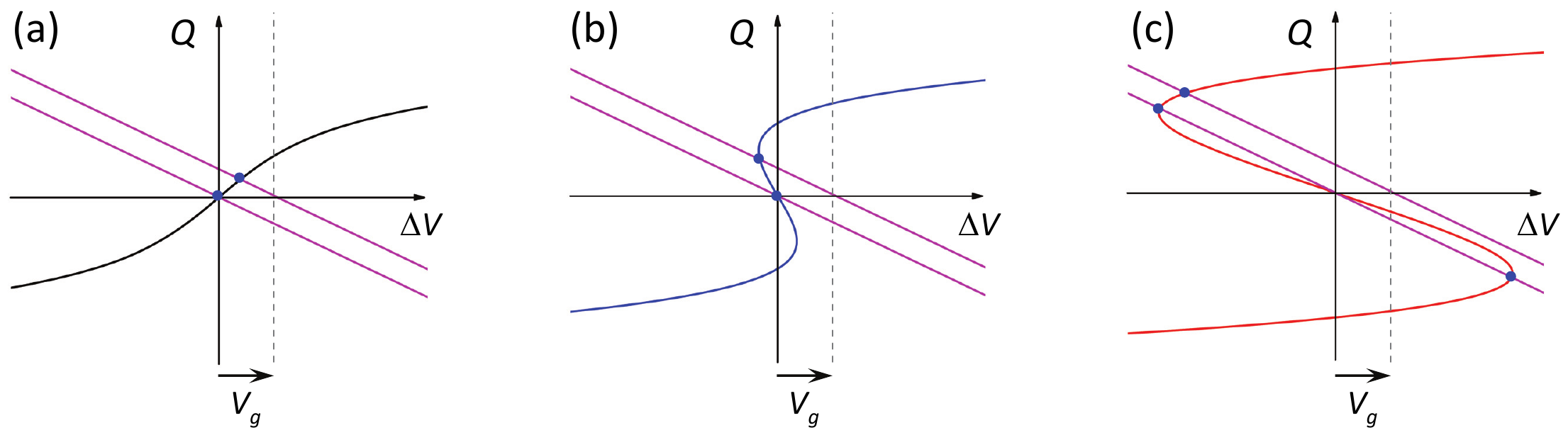}
\caption{
Sequence of possible states in the ferroelectric FET and geometrical determination of the bias point: (a) and (b) paraelectric states, and (c) hysteretic ferroelectric state. In (b) the negative capacitance regime with the step-up conversion of the surface potential is achieved. 
}
\label{geometricFET}
\end{figure*}

The physics behind this negative capacitance regime is associated to depolarizing field effects, i.e., the electric field that accompanies the polarization of the (finite-size) ferroelectric. As a result of this field, there is a shift in ferroelectric transition point and the ferroelectric can operate in its (otherwise unstable) paraelectric state. 
What is more, in such a state, the voltage drop $\Delta V$ through the ferroelectric decreases by increasing the gate voltage. 
This yields the desired amplification, since the changes in the surface potential $V_s$ are then larger than the ones in the gate voltage ($V_g = \Delta V + V_s$). 
This possibility is explained in Fig. \ref{geometricFET}. 
Here we plot the load line $Q = C_s V_s = C_s(V_g- \Delta V)$, where $Q$ and $C_s$ are the charge and the semiconductor capacitance respectively, and the $Q( \Delta V)$ characteristic of the ferrolectric. The slope of this later is always positive if the ferroelectric is well inside its paraelectric phase [Fig. \ref{geometricFET} (a)] and therefore the intersection between this function and the load line shifts towards higher voltages if the gate voltage is increased. 
The ferroelectric then behaves as the conventional oxide in the FET. 
However, when the ferroelectric $Q( \Delta V)$ characteristic adquires its $S$-shaped form, its slope is negative for $\Delta V =0$ [Fig. \ref{geometricFET} (b) and (c)]. This happens below the nominal transition temperature of the ferroelectric (see below). If such a slope is more negative than $-C_s$ [Fig. \ref{geometricFET} (b)], then the intersection with the load line shifts towards lower voltages as the gate voltage increases. This means that the surface potential is enhanced as we explained before, which can associated with a negative-capacitance behavior of the ferroelectric. If the ferroelectric $Q( \Delta V)$ characteristic gets sufficiently flat for low voltages then there appears three points of intersection with the load line [Fig. \ref{geometricFET} (c)], from which only the marked with dots correspond to stable states for the ferroelectric (now in its ferroelectric state). 
The voltage amplification holds until these points shift again towards higher values if the gate voltage is increased as shown in Fig. \ref{geometricFET} (c). This eventually translates into hysteresis loops for gate voltages varying cyclically from positive to negative. 

%%%%%%%%%%%%%%%%%%%%%%%%%%%%%%%%%%%%%%%%%%%%%%%%%
%%%%%%%%%%%%%%%%%%%%%%%%%%%%%%%%%%%%%%%%%%%%%%%%%

%%%%%%%%%%%%%%%%%%%%%%%%%%%%%%%%%%%%%%%%%%%%%%%%%
%%%%%%%%%%%%%%%%%%%%%%%%%%%%%%%%%%%%%%%%%%%%%%%%%

According to these reasonings, the maximum amplification of the gate potential is expected to be limited by the eventual transition into the ferroelectric state. 
Salahuddin and Datta have tacitly assumed that this transition implies the single-domain state in which the spontaneous polarization is uniform through the ferroelectric. 
The actual situation, however, can be far more subtle. 
As a result of the depolarizing field the ferroelectric instability is generally equivalent to the appearance of multi-domain ferroelectricity in which the polarization varies in space, 
with transition temperature shifts rather different from the assumed by Salahuddin and Datta \cite{Bratkovsky09}. 
The aim of this paper is to show that the negative capacitance regime then gets saturated, which puts severe limits to the maximum amplification of the gate voltage that can be obtained in ferroelectric FETs.

To obtain the response of the FET %gate stack 
to the applied voltage we follow the Landau-like approach described in \cite{Bratkovsky09}. On one hand, the behavior of the ferroelectric is described by the equations
\begin{subequations}
\label{eqs.state}\begin{align}
a P + b P^3 - c\nabla^2 P &= -\partial_z V,
\label{eqs.state1}\\
\left( \varepsilon _{\parallel }\left( \partial _{x}^{2}+\partial
_{y}^{2}\right) + \varepsilon _{0}\partial _{z}^{2}\right) V- \partial _{z}P &=0. 
\label{eqs.state2}
\end{align} \end{subequations}
Here $P$ is the distribution of polarization along the ferroelectric $z$-axis and $V$ the electrostatic potential in the ferroelectric. Eq. \eqref{eqs.state1}, which can be derived from a Ginzburg-Landau-Devonshire free energy, represents the constitutive equation for the ferroelectric whose electrostatics, as follows from Maxwell's equations, is eventually determined by Eq. \eqref{eqs.state2}. The instability that gives rise to ferroelectricity is described by the vanishing of the coeffient $a=a'(T-T_c^0)$ as usual, where $T_c^0$ is the nominal transition temperature (in the absence of depolarizing field) %, i.e., with $V=0$), 
while the rest of coefficients are assumed to be positive constants. We thus assume a second-order (continuous) phase transition, which {\it a priori} is the most favorable scenario for the amplification of the FET gate voltage. On the other hand, the semiconductor is assumed to be undoped (or lighly doped) operating within its subthreshold regime as in Ref. \cite{Datta08}. Thus, its mobile carrier density can be neglected and we simply have the equation $\nabla^2 V_s = 0 $ for the electrostatic potential in the semiconductor. At the ferroelectric-semiconductor interface these quantities have to satisfy the electrostatic matching conditions 
$V =V_{s}$ and 
$\varepsilon _{0}\partial _{z}V- P =\varepsilon _{s}\partial _{z}V_{s}$,
where $\varepsilon _{s}$ is the dielectric constant of the semiconductor \cite{note}. 
In addition, we have the boundary conditions 
$V=V_{g}$ at the gate and 
$V_{s}=0$ in the semiconductor beyond its depletion layer (see Fig. \ref{FE-FET}). $V_g$ is assumed to be below the FET threshold voltage in the following. 

\begin{figure}[tb]
\includegraphics[width=.4\textwidth]{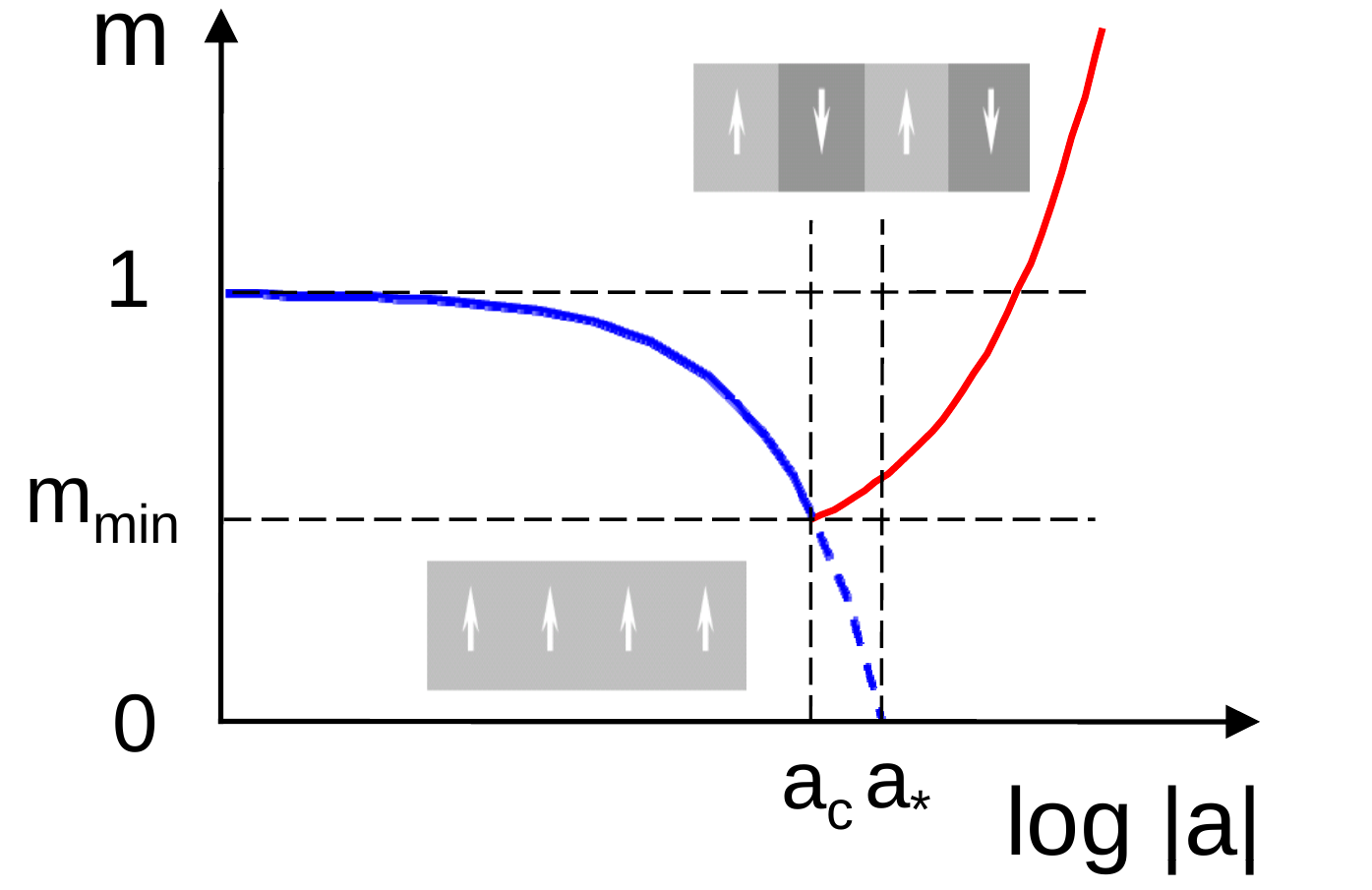}
\caption{
Schematic behavior of the ferroelectric FET body factor $m$ as a function of the control parameter $a = a'(T-T_c^0)$: (blue line) ignoring non-uniform distributions of polarization  (red line) taking into account multi-domain ferroelectricity. Since ferroelectricity is expected in a multi-domain state, $m$ remains finite with its minimum at the corresponding transition point $a_c$. Otherwise it could be downed to zero at the paraelectric $\leftrightarrow$ uniform ferroelectric transition point $a_*$. 
}
\label{body}
\end{figure}

As long as the ferroelectric stays in its paraelectric phase the body factor of the FET is given by
\begin{align}
m= 1 + {\varepsilon_s \over 1 + \varepsilon _{0}a }{l \over w}a. 
\label{bodyfactor_para}\end{align} 
Here $l$ and $w$ represent the thickness of the ferroelectric and the width of the semiconductor depletion layer respectively.
Noting that $\varepsilon _{0}|a| \ll 1$ in the vicinity of the ferroelectric instability one obtains $m \simeq  1 + \varepsilon _{s}{l \over w}a = 1+ {C_s \over C_{FE}}$, where $C_s =\varepsilon_s/w$ and $C_{FE} =1/(al) $, which is the result reported by Salahuddin and Datta \cite{Datta08}. The desired up-conversion of the surface potential is obtained if the body factor is $m < 1$. This is possible if the bare polarization stiffness $a$ gets negative and hence the ferroelectric can act as a negative capacitance ($C_{FE}<0$). 
If depolarizing fields were totally screened, that would mean the instability of the paraelectric phase with respect the spontaneous polarization of the system. 
This polarization, however, is accompained with some depolarizing field in the FET, which produces an increase in the polarization stiffness. In consequence, the ferroelectric can remain in its paraelectric phase even if the bare stiffness is $a < 0$. 
In fact, the ferroelectric can be proven to be stable with respect to uniform distributions of polarization up to the point
$a_* = - {1\over \varepsilon _{0}+\varepsilon_s l/w}$ where the expression \eqref{body} for the body factor would give zero. This point, however, generally does not correspond to the transition point in the ferroelectric FET as we show in the following. 

The actual phase transition point is determined by the point at which the equations \eqref{eqs.state} have their first nontrivial solution ($P\not =0$) for $V_g =0$. This can be found by following the general procedure outlined in \cite{Bratkovsky09}. 
For the FET geometry and the typical values for the depletion width of ligthly doped semiconductors ($w \sim 0.1-1\,\mu$m), 
such a solution corresponds to a polarization wave $P_0(x,z) = p_0 \cos (k_x x) \cos (k_z z) $, with $k_z = {\pi\over 2l}$ and $k_x = {(\varepsilon_{\parallel}c)^{-1/4}  }\big({\pi\over 2l}\big)^{1/2}$, and appears at
\begin{align}
a_c \simeq -\sqrt{ c\over \varepsilon_\parallel}{\pi\over l}.
\label{ac}\end{align} 
The parameters entering in this expression can be estimated as $\varepsilon_\parallel/ \varepsilon _0 \sim 1-100$ and $c \sim d_{at}^2/\varepsilon_0 $, where $d_{at}$ is the characteristic atomic distance ($c\sim 10^{-9}-10^{-11}\,\text{Jm}^3\text{/C}^2$, see e.g. \cite{gradient}). 
In the FET setups $w,l \gg d_{at}$, and consequently $a_c\gg a_*$. So, in fact, much before the paraelectric phase can get unstable with respect to the uniform polarization, the ferroelectric transits into its (multi-domain) ferroelectric phase at $a_c$. At this point, the body factor \eqref{body} takes the value
\begin{align}
m_\text{min}\simeq 1 - 
{\varepsilon_s \over \varepsilon_\parallel}
{\pi \sqrt{\varepsilon_\parallel c}\over w},
\label{mmin}\end{align} 
which we anticipate to be the minimum obtainable in the ferroelectric FET. 
It is worth noting that $m_\text{min}$ depends on material parameters of both the semiconductor and the ferroelectric, but not on the thickness of this latter. The above numbers give $m_\text{min} \sim 0.99 $ only, though this could be further reduced to $\sim 0.7 $ considering the state-of-the-art semiconductor capacitance $C_s=0.1 \text{ F/m}^2$ (which however requires strong doping and therefore is beyond our model).

As we have mentioned, the expression \eqref{bodyfactor_para} for the body factor is valid as long as the ferroelectric stays in its paraelectric phase. That is, for $a \geq a_c$. 
To obtain the corresponding expression for $a \leq a_c$ it is important to take into account that there is a non-zero background polarization in the ferroelectric. 
Close to the transition point such a polarization is well described by the polarization wave $P_0$ found before and, to our purposes, higher harmonics can be neglected. Within this approximation the amplitude of the polarization wave is $p_0 \simeq \pm {4\over 3}\sqrt{|a-a_c|/b} $. Furthermore we express the total polarization as $P_{tot} = P_0(x,z) + \delta P$, where $\delta P$ is due to the applied gate voltage, and linearize the equations with respect to this quantity ($P_{tot}^3 \simeq P_0^3 + 3P_0^2 \delta P$). We then obtain the body factor
\begin{align}
m\simeq 1+ \varepsilon _{s}\frac{l}{w}
\left(a_c + \frac{1}{3}|a-a_c|\right).
\end{align}%
As we see, the body factor in fact increases once the ferroelectric enters in its (multi-domain) ferroelectric phase. This hardening results from the cubic $P^3 $ term that eventually stabilizes the system. Nevertheless, the ferroelectric stays in the negative capacitance regime for some range of temperatures below the transition point. The precise computation of this range requires to go beyond the single harmonic approximation to describe properly the evolution of the ferroelectric ground state, which is beyond the scope of the present work. It is worh mentioning that the negative capacitance regime in the (multi-domain) ferroelectric phase also manifests in the unusual hysteresis loops with negative slopes described in Ref. \cite{Bratkovsky06} to rationalize experimental data on ferroelectric thin films \cite{Kim05}. 
The behavior of the body factor is illustrated in Fig \ref{body} as a function of the control parameter $a$. 
The maximum amplification of the gate voltage corresponds to Eq. \eqref{mmin} at the transition point $a_c$. 
We note that, in order to obtain a significant gain, the semiconductor capacitance should be engineered to be $C_s = {1\over \pi}\sqrt{\epsilon_\parallel \over c} $ per surface area. This design rule is independent of the ferroelectric thickness. Such a thickness simply sets the amplification window that, for practical purposes, has to be tuned about room temperature. In a multi-domain scenario the gradient coefficient $c$ plays a more important role, giving the above design rule in sharp contrast the reported in Ref. \cite{Datta08}.

In conclusion, we have shown that appearance of multi-domain ferroelectricity may substantially limit the maximum voltage amplification expected in ferroelectric field-effect-transistors.

D.J. acknowledges support from 
projects Explora TEC2008-01883-E/TEC and TEC2009-09350
%D.J. acknowledges support from project Explora TEC2008-01883-E/TEC.

%%%%%%%%%%%%%%%%%%%%%%%%%%%%%%%%%%%%%%%%%%%%%%%%%
%%%%%%%%%%%%%%%%%%%%%%%%%%%%%%%%%%%%%%%%%%%%%%%%%


\begin{references}

\bibitem{Roadmap} International Technology Roadmap for Semiconductors. 2009 Edition. www.itrs.net

\bibitem{Datta08} S. Salahuddin and S. Datta, Nano Lett. {\bf 8}, 405 (2008). 

\bibitem{Dawber05} See, e.g., M. Dawber, K. M. Rabe and J. F. Scott, \rmp {\bf 77}, 1083 (2005), and the references therein.

%\bibitem{Salvatore08} G.A. Salvatore, D. Bouvet and A.M. Ionescu
%%, Demonstration of Subthrehold Swing Smaller Than 60mV/decade in Fe-FET with P(VDF-TrFE)/SiO2 %Gate Stack, 
%IEEE International Electron Devices Meeting 2008.

\bibitem{Bratkovsky09} 
See, e.g., A.M. Bratkovsky and A.P. Levanyuk, J. Comput. Theor. Nanosci. {\bf 6}, 465 (2009); arXiv:0801.1669, and the references therein.


\bibitem{note} Experimentally, suitable ferroelectric-semiconductor interfaces are
difficult to process and buffer layers are frequently needed to avoid e.g.
interdifussion problems. In addition, different especies such as SiO$_x$ may
appear at the interface. This can be seen as an additional capacitance $C_{int}$ in
series with the semiconductor capacitance $C_s$. Its influence on the body factor
can be easily taken into account by replacing $C_s = \epsilon_s/w$ in Eqs. (2), (4) and
(5) by there resulting effective capacitance $C_s C_{int}/(C_s+C_{int})$.


\bibitem{gradient}
W.Y. Shih, W.-H. Shih and I.A. Aksay, \prb {\bf 50}, 15575, (1994).
O.G. Vendik, S.P. Zubko and L.T. Ter-Martirosayn, Appl. Phys. Lett. {\bf 73}, 37 (1998);
A.G. Zembilgotov, N.A. Pertsev, H. Kohlsted and R. Waser, J. Appl. Phys. {\bf 91} 2247 (2003);

\bibitem{Bratkovsky06} 
A.M. Bratkovsky and A.P. Levanyuk, \apl {\bf 89}, 253108 (2006).

\bibitem{Kim05} D. J. Kim {\it et al.}, Phys. Rev. Lett. 95, 237602 (2005); Y. S. Kim {\it et al.} Appl. Phys. Lett. {\bf 86}, 102907 (2005); Y. S. Kim {\it et al.}, Appl. Phys. Lett. {\bf 88}, 072909 (2006).

\end{references}
\end{document}